\documentclass[aps,preprintnumbers,amsmath,nofootinbib]{revtex4}
\usepackage{amssymb}
\usepackage{graphicx}
\usepackage{dcolumn}
\usepackage{bm}
\usepackage{footnpag}

\begin{document}
\title
{
Energy dependence of femtoscopy properties of pion source in
nuclear collisions}

\author{V.A. Okorokov} \email{VAOkorokov@mephi.ru; okorokov@bnl.gov}
\affiliation{National Research Nuclear University "MEPhI",
Kashirskoe Shosse 31, 115409 Moscow, Russia}

\date{\today}

\begin{abstract}
In the paper energy dependence of femtoscopy characteristics of
pion emission region at freeze-out is investigated for collisions
of various ions and for all experimentally available energies. For
the first time the normalized values of radii and volume of source
are used for energy dependence. This approach allows us to expand
the set of interaction types, in particular, on non-symmetrical
nucleus-nucleus collisions which can be studied in the framework
of common approach. There are no the sharp changing of femtoscopic
parameter values, in particular,
$R_{\mbox{\scriptsize{o}}}/R_{\mbox{\scriptsize{s}}}$ with
increasing of $\sqrt{\smash[b]{s_{\footnotesize{NN}}}}$ which were
predicted by some phenomenological models as signature of first
order phase transition in strongly interacting matter. The
generalized parameterization for femtoscopic correlation function
is suggested.

\end{abstract}

\maketitle

\section{Introduction}\label{intro}
At the present time the presence of relationship between the
geometry of 4-dimensional emission region of secondary particles
and dynamics of final state creation is established reliably for
different types of interactions. By the definition (see, for
example,
\cite{Weiner-book-1-2000,Lednicky-PAN-67-72-2004,Okorokov-UchPosob-2009})
the field of research investigated the pair correlations for
secondary particles (both the identical and the non-identical)
with small relative momentum / velocity is called "correlation
femtoscopy". The momentum quantum-statistical correlations
(GGLP-effect) was observed for the first time in
$\bar{\mbox{p}}\mbox{p}$ annihilation for identical charged pions
\cite{Goldhaber-PR-120-300-1960}. The bases of correlation
femtoscopy are described in detail elsewhere (see, for example,
\cite{Weiner-book-1-2000,Okorokov-UchPosob-2009}). The discussion
below is focused on correlations in pairs of identical charged
pions with small relative momenta --
HBT-interferometry\footnote{The detail explanation is shown in
\cite{Lednicky-PAN-67-72-2004} for terminology and relationship
between momentum GGLP correlations in physics of fundamental
interactions and space-time correlations of photons -- HBT-effect
\cite{Hanbury-Brown-PhylMag-45-663-1954}. In accordance with the
field of research of the paper and most popular terminology in
corresponding literature the method of correlation femtoscopy for
identical particles (pions) is called "HBT-interferometry" or
briefly "HBT" also in this paper below.} -- in nucleus-nucleus
collisions. Originally the GGLP-effect was interpreted in the
framework of formalism of wave function
\cite{Goldhaber-PR-120-300-1960,Kopylov-YaF-15-392-1972}, but the
more general approach is based on the quantum field theory
\cite{Weiner-book-1-2000}. This approach was developed and used
for the first time in quantum optics \cite{Weiner-book-1-2000}.

The space-time characteristics for emission region of secondary
particles created in (heavy) ion collisions are important for
study of deconfinement state of strongly interacting matter --
strong-coupling quark-gluon plasma (sQGP): the significant
increasing of emission duration was predicted as compulsory
signature of the first order phase transition in strongly
interacting matter from hadronic phase to quark-gluon one
\cite{Pratt-PRD-33-1314-1986,Bertsch-NPA-498-173c-1989}. This
effect should be shown experimentally as strong difference of
emission region in transverse plane from azimuthal-symmetric
shape. The correlation femtoscopy allows us to investigate the
geometry of source at kinetic freeze-out, i.e. at late stage of
space-time evolution of final state at transition from strongly
coupling system to weakly interacting ensemble of secondary
particles. Therefore the study of $\mbox{A}+\mbox{A}$ collisions
in wide energy domain by correlation femtoscopy seems important
for better understanding both of equation of state (EOS) of
strongly interacting matter and general dynamic features of soft
processes.

It should be emphasized that there is deep relationship between
collective effects at various stages of space-time evolution of
strong interaction processes. For example, the measurements of
length scales and chaoticity of source with help of correlation
femtoscopy can be used for estimation of multiplicity of hadron
jets without application of specific algorithms for jet
identification \cite{Alexander-arXiv-1202.3575}, for study of
differences between quark and gluon jets
\cite{Zheng-arXiv-1302.4511-2013}. The correlation femtoscopy
seems the promising tool for investigation of fundamental discrete
symmetries \cite{Vance-PRL-81-2205-1998} and complex structure of
quantum chromodynamic (QCD) vacuum. Moreover, the geometry of
emission region is important for physics of cosmic rays and for
search for signatures of physics beyond of Standard Model (SM)
\cite{Srivastava-arXiv-1201.2380-2012}. Therefore the studies in
the field of the correlation femtoscopy have a inter-subject
character.

The paper is organized as follows. In Sec.\,2, definitions of main
observables for correlation femtoscopy are described. The
normalized characteristics for geometry of emission region are
defined. The Sec.\,3 devotes discussion of experimental energy
dependence for the femtoscopic parameters of secondary particle
source at freeze-out in various ion collisions. In Sec.\,4, the
general case of L\'{e}vy stable distribution is studied for
correlation function. Some final remarks and conclusions are
presented in Sec.\,5.

\section{Method and variables}\label{sec:2}
In general two-particle correlation function (CF) for secondary
particles of type $j$ with 4-momenta $p_{\,1}, p_{\,2}$ is defined
as follows \cite{Pratt-PRL-53-1219-1984}:
\begin{equation}
C_{2}^{j}(p_{\,1},p_{\,2}) = \frac{\textstyle
\mathbf{P}_{2}^{j}(p_{\,1},p_{\,2})}{\textstyle
\mathbf{P}_{1}^{j}(p_{\,1})\mathbf{P}_{1}^{j}(p_{\,2})} =
\sigma_{\mbox{\scriptsize{in}}}\frac{\textstyle
d^{\,2}\sigma^{j}(p_{\,1},p_{\,2}) / dp_{\,1}dp_{\,2}}{\textstyle
d\sigma^{j} / dp_{\,1} \times d\sigma^{j} / dp_{\,2}},
\label{eq:2.1}
\end{equation}
where $\mathbf{P}_{(1)\,2}^{j}$ is (one)two-particle inclusive
distribution density, $\sigma_{\mbox{\scriptsize{in}}}$ -- the
total inelastic cross section for interaction under study. There
is the following relationship between (\ref{eq:2.1}) and
normalized cumulant CF \cite{DeVolf-UFN-163-3-1993}
\begin{equation}
\mathbf{K}_{2}^{j}(p_{\,1},p_{\,2})=C_{2}^{j}(p_{\,1},p_{\,2})-1.\label{eq:2.2}
\end{equation}
Functions (\ref{eq:2.1}) and (\ref{eq:2.2}) are studied depending
on relative 4-momentum $q \equiv (q^{0},\vec{q})=p_{\,1}-p_{\,2}$
and average 4-momentum of particles in pair $K \equiv
(K^{0},\vec{K})=(p_{\,1}+p_{\,2})/2$ (pair 4-momentum).
Phenomenological multidimensional parameterization for CF
(\ref{eq:2.1}) for standard simplest case can be written as (see,
for example, \cite{Okorokov-UchPosob-2009})\footnote{For
discussion below the index for particle type is omitted for
simplicity.}:
\begin{equation}
C_{2}^{\,\mbox{\scriptsize{ph}}}(q,K) \propto 1+
\lambda(K)\mathbf{K}_{2}^{\mbox{\scriptsize{ph}}}({\bf A}),~~~
\mathbf{K}_{2}^{\mbox{\scriptsize{ph}}}({\bf
A})=\prod\limits_{i,j=1}^{3}
\mathbf{K}_{2}^{\mbox{\scriptsize{ph}}}(A_{ij})=
\exp\biggl(-\sum\limits_{i,j=1}^{3}q_{i}R_{ij}^{\,2}q_{j}\biggr).
\label{eq:2.3}
\end{equation}
Here ${\bf A} \equiv \vec{q}\,{\bf R}^{2}\vec{q}^{\,T}$ and ${\bf
R}^{2}$ are the matrices $3 \times 3$, $\vec{q}^{\,T}$ --
transposed vector $\vec{q}$, $\forall~ i,j:
R^{\,2}_{ij}=R^{\,2}_{ji}, R^{\,2}_{ii} \equiv R^{\,2}_{i}$, where
$R_{i}=R_{i}(K)$ are parameters derived by HBT method and
characterized the linear scales of source part which can be
studied at fixed $K$, i.e. homogeneity region
\cite{Sinyukov-NATOSeries-346-309-1995}; the products are taken on
space components of vectors, $\lambda(K)=\mathbf{K}_{2}(0,K), 0
\leq \lambda \leq 1$ is the parameter which characterize the
degree of source chaoticity\footnote{It should be emphasized that
this title of $\lambda$ is the historical and can be used with
some carefulness because it is valid for fully chaotic source
without any other features of dynamics (contribution of long-lived
resonances etc.) and experiment. The set of effect contributions
included in $\lambda$ depends on certain investigation. As usual
the $\lambda$ takes into account the partial coherence of source
and long-lived resonance decays in theoretical studies. Within the
framework of phenomenological and experimental investigations the
$\lambda$ includes both the effects indicated above and
contributions of weak decays and particle misidentification. The
effects of final state interactions (see below) can be accounted
in $\lambda$ in some rare case. The separation of influence of
source coherence on $\lambda$ from contribution of other effects
can be made with help of three-pion correlation only. These
correlations suppress contributions from long-lived resonance
decays and particle misidentification significantly
\cite{Heinz-PRC-56-426-1997}. The physical analysis of
three-particle correlations allows us to define the true degree of
source coherence and, as consequence, influence just of the
coherence on the $\lambda$ values estimated with help the study of
$C_{2}(q,K)$.}. Taking into account the hypothesis of cylindrical
symmetry of source the volume of homogeneity region was derived as
follows \cite{Okorokov-ISHEPP-101-2006}
\begin{equation}
V=(2\pi)^{3/2}\prod\limits_{i=1}^{3}R_{i}. \label{eq:2.4}
\end{equation}

The experimental correlation function is constructed as follows
\cite{Okorokov-UchPosob-2009}
\begin{equation}
C_{2}^{\mbox{\scriptsize{E}}}(q,K)=\zeta(q,K)
D_{\mbox{\footnotesize{E}}}(q,K)
D_{\mbox{\footnotesize{B}}}^{-1}(q,K), \label{eq:2.5}
\end{equation}
where $D_{\mbox{\footnotesize{E}}}(q,K)$ is the pair distribution
for particles measured in the same event,
$D_{\mbox{\footnotesize{B}}}(q,K)$ -- background distribution --
distribution for pairs of particles from different events. In an
ideal case the background distribution is the same as
$D_{\mbox{\footnotesize{E}}}(q,K)$ with exception of presence of
quantum-statistical correlations in the last case. The additional
factor $\zeta(q,K)$ takes formally into account all possible
corrections. It was shown that the ratio (\ref{eq:2.5}) is
sensitive to the space-time extension of emission region
\cite{Lisa-AnnRevNuclPartSci-55-357-2005}.

The space component of pair 4-momentum ($\vec{K}$) is decomposed
on longitudinal
$k_{\parallel}=(p_{\,\parallel,1}+p_{\,\parallel,2})/2$ and
transverse
$\vec{k}_{\perp}=(\vec{p}_{\perp,1}+\vec{p}_{\perp,2})/2$ parts of
pair momentum. There are several version for decomposition of $q$
\cite{Pratt-PRD-33-1314-1986,Bertsch-PRC-37-1896-1988,Yano-PLB-78-556-1978,Csorgo-HIP-15-1-2002}.
In the paper decomposition of Pratt -- Bertsch
\cite{Pratt-PRD-33-1314-1986,Bertsch-PRC-37-1896-1988} is used in
which the $\vec{q}$ is resolved  into longitudinal component
directed along the beam axis, $q_{\mbox{\scriptsize{l}}}$, outward
component directed parallel to the pair transverse momentum,
$q_{\mbox{\scriptsize{o}}}$, and a sideward component directed
perpendicular to those two, $q_{\mbox{\scriptsize{s}}}$. For
identical particle pairs the longitudinal co-moving system (LCMS)
frame is chosen as the reference frame. The LCMS moves together
with pair in longitudinal direction, thus the $k_{\parallel}=0$ in
this frame. In general case for nuclear beam collisions
$\mbox{A}_{1} + \mbox{A}_{2}$ one can write for length scales of
homogeneity region
$R_{i}=f(\sqrt{\smash[b]{s_{\footnotesize{NN}}}}, \mbox{A}_{1},
\mbox{A}_{2}, |\vec{b}|, \phi, y, |\vec{k}_{\perp}|, m)$, where
$i=\mbox{l,\,o,\,s}$, $\vec{b}$ is the impact parameter vector,
$\phi$, $y$ -- azimuthal angle and rapidity. The main part of
measurements for correlation femtoscopy in field of heavy ion
interactions was made for pairs of identical charged pions in
central collisions. This allows us to simplify the theoretical
formalism significantly due to azimuthal symmetry
\cite{Lisa-AnnRevNuclPartSci-55-357-2005,Heinz-PRC-66-044903-2002}
and to reach the maximum for energy density and linear sizes of
emission region.

\begin{figure*}
\includegraphics[width=15.5cm,height=17.0cm]{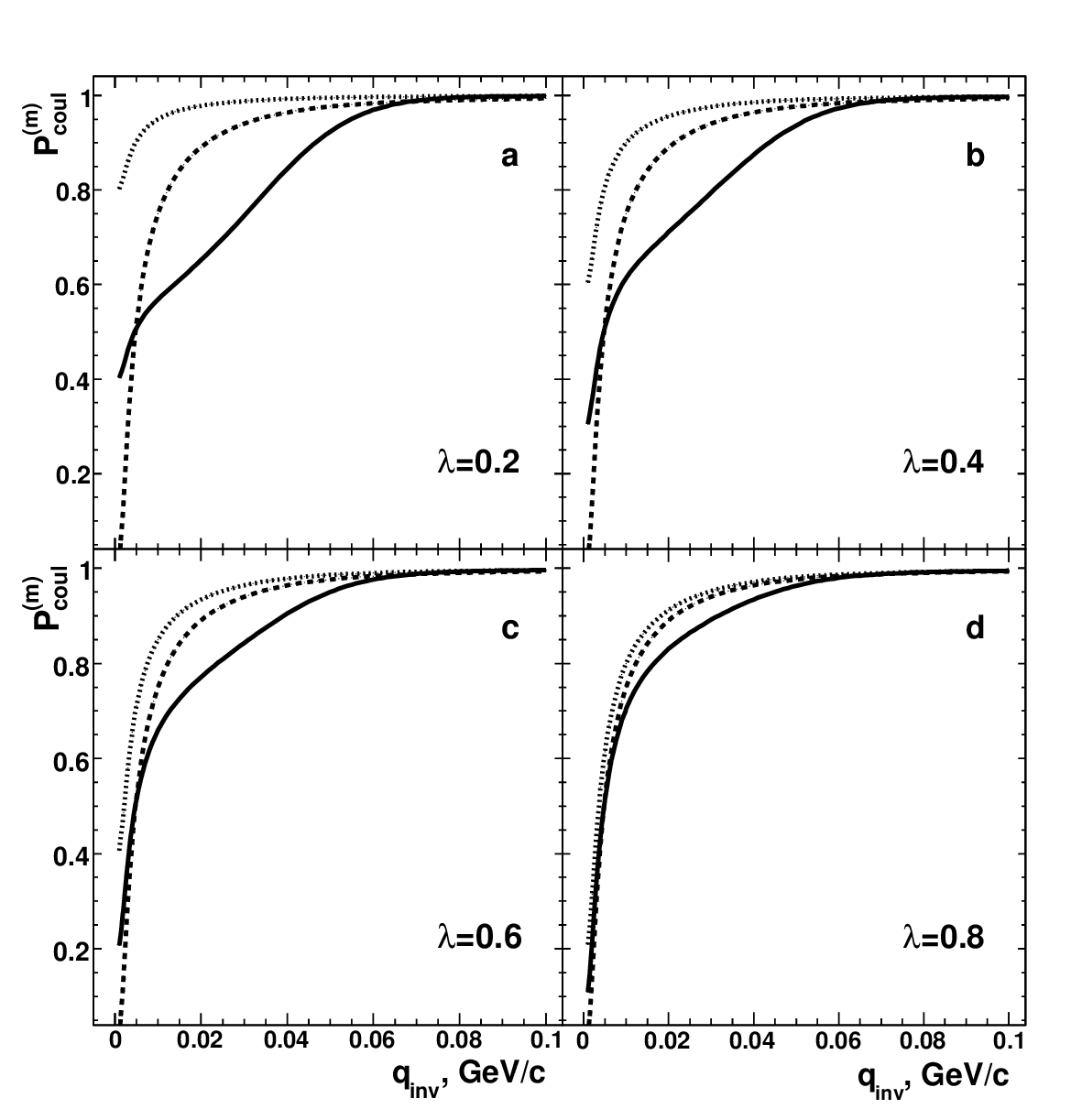}
\vspace*{8pt} \caption{The values of correction on Coulomb FSI
$P_{\mbox{\scriptsize{coul}}}^{(m)}$ depends on
$q_{\mbox{\scriptsize{\,inv}}}$ for different values of $\lambda$,
where $m$ is the number of procedure for definition of Coulomb
correction. The corrections are calculated for static
spherically-symmetrical Gaussian source with $R=6$ fm. The dashed
line corresponds for $m=1$, dotted line -- $m=2$, solid line --
$m=3$.} \label{fig:0}
\end{figure*}

Conservation laws and interactions of particles in final state
(FSI) influence on quantum-statistical correlations, moreover just
the FSI is most important for nuclear collisions
\cite{Wiedemann-PR-319-145-1999}. The methods to take into account
of FSI in general case are described in
\cite{Gyulassy-PRC-20-2267-1979}. For the pairs of charged hadrons
the main contribution is Coulomb FSI but correction due to strong
interactions influences much weaker in the correlation peak domain
$|\vec{q}| \leq 0.1$ GeV/$c$ at the same time
\cite{Wiedemann-PR-319-145-1999}. As known the Coulomb repulsion
in the pairs of same-sign charged particles leads to decreasing of
amount of real pairs at small $q$ and consequent decreasing of
peak amplitude of CF. Therefore the function
\begin{equation}
C_{2}^{\mbox{\scriptsize{E}}}(q,K)=
\bigl[D_{\mbox{\footnotesize{E}}}(q,K)
D_{\mbox{\footnotesize{B}}}^{-1}(q,K)\bigr]
P_{\mbox{\footnotesize{coul}}}^{-1}(q). \label{eq:2.6}
\end{equation}
is called corrected experimental CF in the field of correlation
femtoscopy, where $P_{\mbox{\footnotesize{coul}}}(q)$ is the
correction on the Coulomb FSI. It would be noted that
$P_{\mbox{\footnotesize{coul}}}(q)$ is introduced either in
phenomenological parameterization (\ref{eq:2.3}) or in
experimental CF. In the first case the
$C_{2}^{\,\mbox{\scriptsize{ph}}}(q,K)$ is multiplied on
$P_{\mbox{\footnotesize{coul}}}(q)$ but in the last case the
experimental CF is divided on the correction on the Coulomb FSI
and the equation (\ref{eq:2.6}) is derived. The main procedures
for accounting for Coulomb FSI are described in
\cite{Pratt-PRD-33-72-1986,Ahle-PRC-66-054906-2002,Bowler-PLB-270-69-1991}.
The standard procedure is suggested the iteration procedure for
calculation of Coulomb correction
$P^{(1)}_{\mbox{\footnotesize{coul}}}(q)$ for extended source. The
model approach for the source is the static
spherically-symmetrical Gaussian source with fixed radius $R$ in
the rest frame of pair \cite{Pratt-PRD-33-72-1986}. Historically
this procedure was suggested as first and was used in the many
experiments, in particular, for the first HBT study of
$\mbox{Au+Au}$ collision at RHIC energy
$\sqrt{s_{\footnotesize{NN}}}=130$ GeV
\cite{Adler-PRL-87-082301-2001}. However this procedure some
overestimates the value of correction on the Coulomb FSI due to
suggestion that all pairs in $D_{\mbox{\footnotesize{B}}}(q,K)$
are primary and should be corrected \cite{Pratt-PRD-33-72-1986}.
This aspect is taken into account in framework of the second
procedure \cite{Ahle-PRC-66-054906-2002} by excluding the pairs
which are formed by pions from resonance decays and participate in
the Coulomb interaction at the same time. The such exclusion leads
to some attenuation of Coulomb correction and the following
relation is derived for $P_{\mbox{\footnotesize{coul}}}(q)$ in the
framework of second procedure
$P^{(2)}_{\mbox{\footnotesize{coul}}}(q)=(1-f_{(2)})+f_{(2)}P^{(1)}_{\mbox{\footnotesize{coul}}}
(q),$ where $f_{(2)}$ is the fraction of primary pions and $0 \leq
f_{(2)} \leq 1$. This procedure is called the dilution procedure
respectively. In the framework of third procedure it is suggested
that only pairs followed by Bose\,--\,Einstein statistics
participate in the Coulomb FSI \cite{Bowler-PLB-270-69-1991}.
These pairs are formed by the particles which are close to each
other in the center of mass of pair system. In the third procedure
it is suggested that there are no misidentified pairs and
corresponding correction is defined as follows:
$P^{(3)}_{\mbox{\footnotesize{coul}}}(q)=(1-f_{(3)})[1+
\mathbf{K}_{2}^{\mbox{\scriptsize{ph}}}({\bf A})]^{-1} +f_{(3)}
P^{(1)}_{\mbox{\footnotesize{coul}}}(q),$ where $f_{(3)}$ is the
fraction of pairs followed by Bose\,--\,Einstein statistics and $0
\leq f_{(3)} \leq 1$. In the equation above the first term
corresponds to the pairs which do not participate in the Coulomb
FSI, the second term -- to the pairs which follows by
Bose\,--\,Einstein statistics and participate in the Coulomb FSI
at the same time. The choice $\forall~m=2,3: f_{(m)}=\lambda$
seems reasonable for fully chaotic source and the specific
physical meaning of the $\lambda$ corresponds to the certain
procedure for Coulomb FSI. As seen from the relations for
$P^{(m)}_{\mbox{\footnotesize{coul}}}(q)$ at $m=2,3$ all procedure
for accounting of Coulomb FSI are identical at $\lambda=1$. Fig.
\ref{fig:0} shows the dependence of various Coulomb corrections
$P_{\mbox{\scriptsize{coul}}}^{(m)}$ on
$q_{\mbox{\scriptsize{\,inv}}}=\sqrt{-q^{2}}$, the corrections
were obtained for model of static spherically-symmetrical Gaussian
source with $R=6$ fm. As expected the difference between various
procedures decrease with $\lambda$ increasing. Therefore the
following general equation can be written for phenomenological
parameterization of CF with taking into account all forms of
corrections on Coulomb FSI under consideration
\begin{equation}
C_{2,(m)}^{\mbox{\scriptsize{ph}}}(q,K)=\epsilon
P_{\mbox{\footnotesize{coul}}}^{(m)}(q)
\bigl[\epsilon^{-1}+\mathbf{K}_{2}^{\mbox{\scriptsize{ph}}}({\bf
A})\bigr], ~~\epsilon=\left\{
\begin{array}{ll}
\lambda, & \mbox{at}~m=1,2;\\
1, & \mbox{at}~m=3.
\end{array}
\right.\label{eq:2.7}
\end{equation}

Because of complex dynamics and space-time structure of emission
region in $\mbox{A}+\mbox{A}$ interactions at both intermediate
and high $\sqrt{\smash[b]{s_{\footnotesize{NN}}}}$ the difference
is possible between pair ensembles which participate in
Bose\,--\,Einstein correlations and Coulomb FSI. Therefore the
third procedure for calculation of Coulomb correction seems most
adequate for study of heavy ion collisions. The correction
$P_{\mbox{\scriptsize{coul}}}^{(3)}$ was used for femtoscopic
analysis in energy domain from SPS to LHC in various experiments
(see below Sec.\,3).

In the paper the following set of femtoscopic observables
$\mathcal{G} \equiv \{\mathcal{G}_{i}\}_{i=1}^{5}=\{\lambda,
R_{\mbox{\scriptsize{s}}}, R_{\mbox{\scriptsize{o}}},
R_{\mbox{\scriptsize{l}}},V\}$ is under consideration. This set of
parameters characterizes the chaoticity of source and its
4-dimensional geometry at freeze-out stage completely. Moreover in
the paper the using of normalized values of HBT-radii and volume
of homogeneity region is suggested in order to extend the set of
types of collisions which can be studied within the framework of
general approach. The normalized femtoscopy parameters
$\mathcal{G}_{i}$, $i=2-5$ are calculated as follows:
\begin{equation}
R_{i}^{n}=R_{i}/R_{\mbox{\scriptsize{A}}},~
i=\mbox{s,o,l};~~~V^{n}=V/V_{\mbox{\scriptsize{A}}}.\label{eq:2.8}
\end{equation}
Here $R_{\mbox{\scriptsize{A}}}=r_{0}A^{1/3},
V_{\mbox{\scriptsize{A}}}=4\pi R^{3}_{\mbox{\scriptsize{A}}}/3$
are radius and volume of spherically-symmetric nucleus,
$r_{0}=(1.25 \pm 0.05)$ fm \cite{Mukhin-book-1983}. In the case of
non-symmetric nucleus-nucleus collisions the factors for
normalization in (\ref{eq:2.8}) is defined by $\langle
R_{\mbox{\scriptsize{A}}}\rangle=0.5(R^{1}_{\mbox{\scriptsize{A}}}+R^{2}_{\mbox{\scriptsize{A}}})$
is the arithmetical mean value of radii of colliding ions, where
$\forall\,i:~R^{i}_{\mbox{\scriptsize{A}}}$ is calculated based on
the equation above for spherically-symmetric nucleus.

\section{Energy dependence for the femtoscopic parameters}\label{sec:3}
The investigation of energy dependence of femtoscopic parameters
from set $\mathcal{G}$ seems important, in particular, for search
of creation of deconfinement state of strongly interacting matter
(sQGP) in $\mbox{A+A}$ collisions and for study of corresponding
phase transitions. In accordance to some theoretical predictions
the methods of correlation femtoscopy allow us to search for
qualitatively new physical effects in RHIC energy domain.
Therefore the comparison of femtoscopy results at various initial
energies is important for energy range as wide as possible. The
dependencies of $\{\mathcal{G}_{i}\}_{i=1}^{4}$ and
$R_{\mbox{\scriptsize{o}}}/R_{\mbox{\scriptsize{s}}}$ on
$\sqrt{\smash[b]{s_{\footnotesize{NN}}}}$ for secondary pions were
shown elsewhere, for example,
\cite{Adler-PRL-87-082301-2001,Okorokov-ISMD-137-2002,Okorokov-PRC-71-044906-2005}.
The corresponding dependencies for charged kaons were discussed in
\cite{Okorokov-ISHEPP-101-2006} for the first time. It would be
noted that some new experimental results were obtained for the
last years, in particular, the range of collision energy was
extended in the TeV-region for secondary pions. Therefore in the
paper the dependencies of set $\mathcal{G}$ of femtoscopic
parameters on $\sqrt{\smash[b]{s_{\footnotesize{NN}}}}$ are
studied based on the all available experimental results which were
obtained in the framework of approach for gaussian shape of
correlation function.

\begin{figure*}
\includegraphics[width=15.5cm,height=17.0cm]{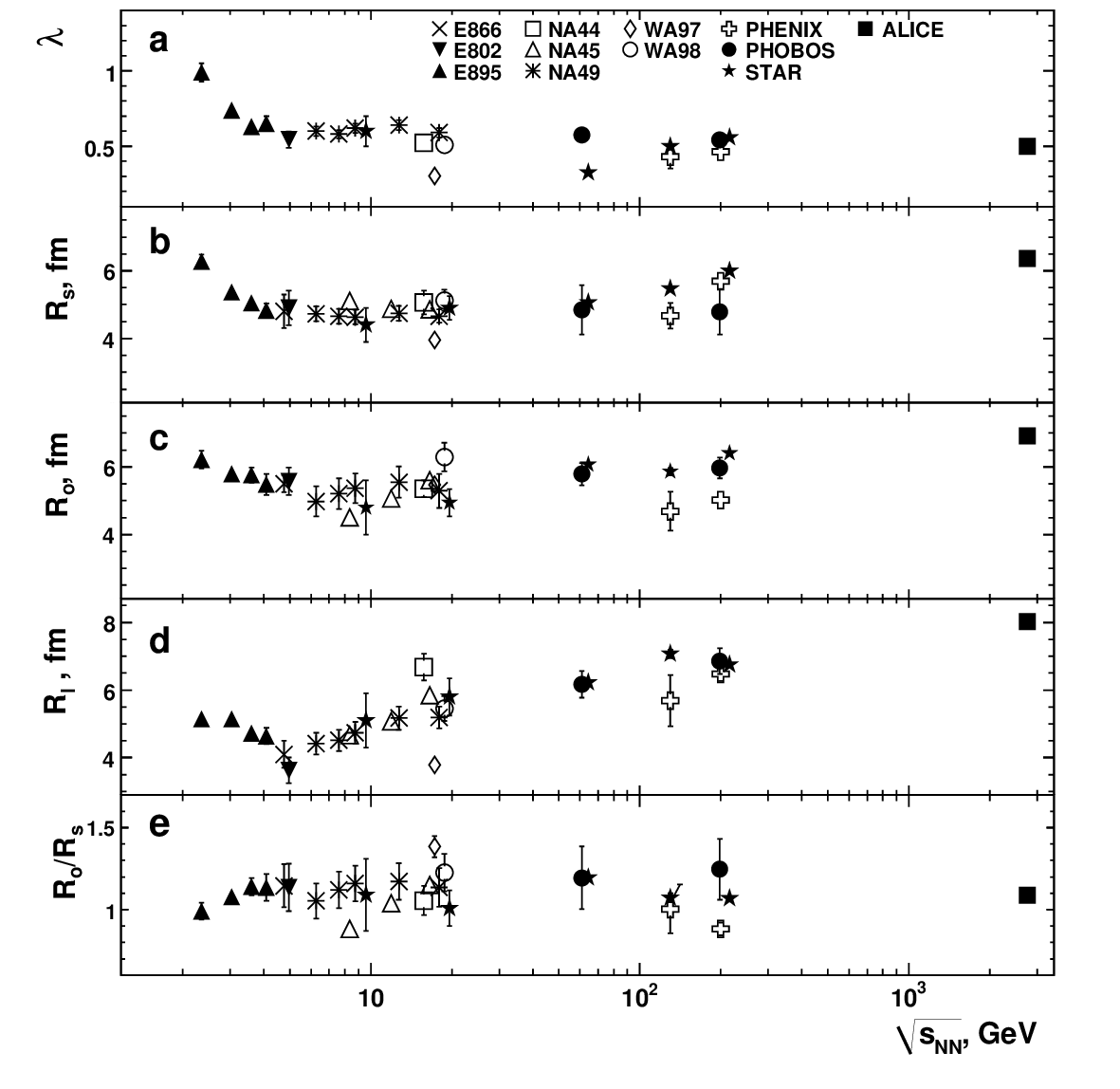}
\vspace*{8pt} \caption{Dependencies of chaoticity parameter (a),
HBT-radii (b -- d) and ratio
$R_{\mbox{\scriptsize{o}}}/R_{\mbox{\scriptsize{s}}}$ (e) on
initial energy for central heavy ion $\mbox{Au+Au, Au+Pb, Pb+Pb}$
collisions at midrapidity and $\langle k_{\perp}\rangle \simeq
0.2$ GeV/$c$
\cite{Adler-PRL-87-082301-2001,Okorokov-PRC-71-044906-2005,PHENIX-PRL-93-152302-2004,
Ahle-PRC-66-054906-2002,Soltz-NPA-661-439c-1999,Bearden-PRC-58-1656-1998}.
Experimental results are demonstrated for pairs of $\pi^{-}$
mesons (in the case of ALICE -- for $\pi^{\pm}\pi^{\pm}$ pairs)
and for standard Coulomb correction
$P_{\mbox{\scriptsize{C}}}^{(1)}(q)$ (in cases of ALICE, NA44,
NA45, PHOBOS and STAR at
$\sqrt{\smash[b]{s_{\footnotesize{NN}}}}=62.4$ GeV -- for
correction $P_{\mbox{\scriptsize{C}}}^{(3)}$). Statistical errors
are shown (for NA44 -- total uncertainties).} \label{fig:1}
\end{figure*}

Dependencies of femtoscopic parameters
$\mathcal{G}_{i}(\sqrt{\smash[b]{s_{\footnotesize{NN}}}})$,
$i=1-4$ and
$R_{\mbox{\scriptsize{o}}}/R_{\mbox{\scriptsize{s}}}(\sqrt{\smash[b]{s_{\footnotesize{NN}}}})$
are shown in Figs. \ref{fig:1}a -- d and Fig. \ref{fig:1}e
respectively. The experimental results have been obtained for
identical charged pion pairs with low $k_{\perp}$ and midrapidity
in (quasi)symmetric heavy ion collisions. For correct comparison
with results from previous measurement at intermediate energies
the values of femtoscopic parameters at RHIC are shown for
standard gaussian approximation of
$C_{2}^{\mbox{\scriptsize{E}}}(q,K)$ and Coulomb correction
$P_{\mbox{\scriptsize{coul}}}^{(1)}(q)$. The more careful
correction of correlation function on Coulomb final state
interaction with help of $P_{\mbox{\scriptsize{coul}}}^{(3)}(q)$
leads to some decreasing of $\lambda, R_{\mbox{\scriptsize{s}}}$
and smaller changing of other parameters from $\mathcal{G}$ for
$\mbox{Au+Au}$ collisions at
$\sqrt{\smash[b]{s_{\footnotesize{NN}}}}=200$ GeV in comparison
with standard procedure. These features are observed both in STAR
\cite{Okorokov-PRC-71-044906-2005} and in PHENIX
\cite{PHENIX-PRL-93-152302-2004} experiments at RHIC. Therefore
accounting for results at intermediate energies it can be
suggested that correction type on Coulomb FSI does not depend on
general trends of energy dependencies of femtoscopic parameters in
energy domain, at least $\sqrt{\smash[b]{s_{\footnotesize{NN}}}}
\simeq 17 - 200$ GeV. As seen there is increasing of HBT radii
(Figs. \ref{fig:1}b -- d) at growth of collision energy from
$\sqrt{\smash[b]{s_{\footnotesize{NN}}}} \sim 20$ GeV up to
maximum available LHC energy
$\sqrt{\smash[b]{s_{\footnotesize{NN}}}}=2.76$ TeV. The chaoticity
parameter $\lambda$ shows the weak changing at
$\sqrt{\smash[b]{s_{\footnotesize{NN}}}} > 4$ GeV (Fig.
\ref{fig:1}e). One can see that pion source is far from fully
chaotic at LHC energy ($\lambda \approx 0.5$). Taking into account
the results at RHIC and type of secondary particles under study
(pions) one can suggest that decays of various resonance states
influences on the chaoticity parameter even at LHC energy and
leads to amplification of coherence of source. The study of
multipion correlations is necessary for more definite physical
conclusion. There is no significant increasing of ratio
$R_{\mbox{\scriptsize{o}}}/R_{\mbox{\scriptsize{s}}}$ in all
experimentally available energy domain (Fig. \ref{fig:1}e) which
was predicted in the framework of ideal hydrodynamics for first
order transition from hadronic to quark-gluon matter. Therefore it
should be emphasized that one of the possible signatures of first
order phase transition to the deconfinement state of strongly
interacting matter is absent for soft observables in wide energy
range.

\begin{figure*}
\includegraphics[width=15.5cm,height=17.0cm]{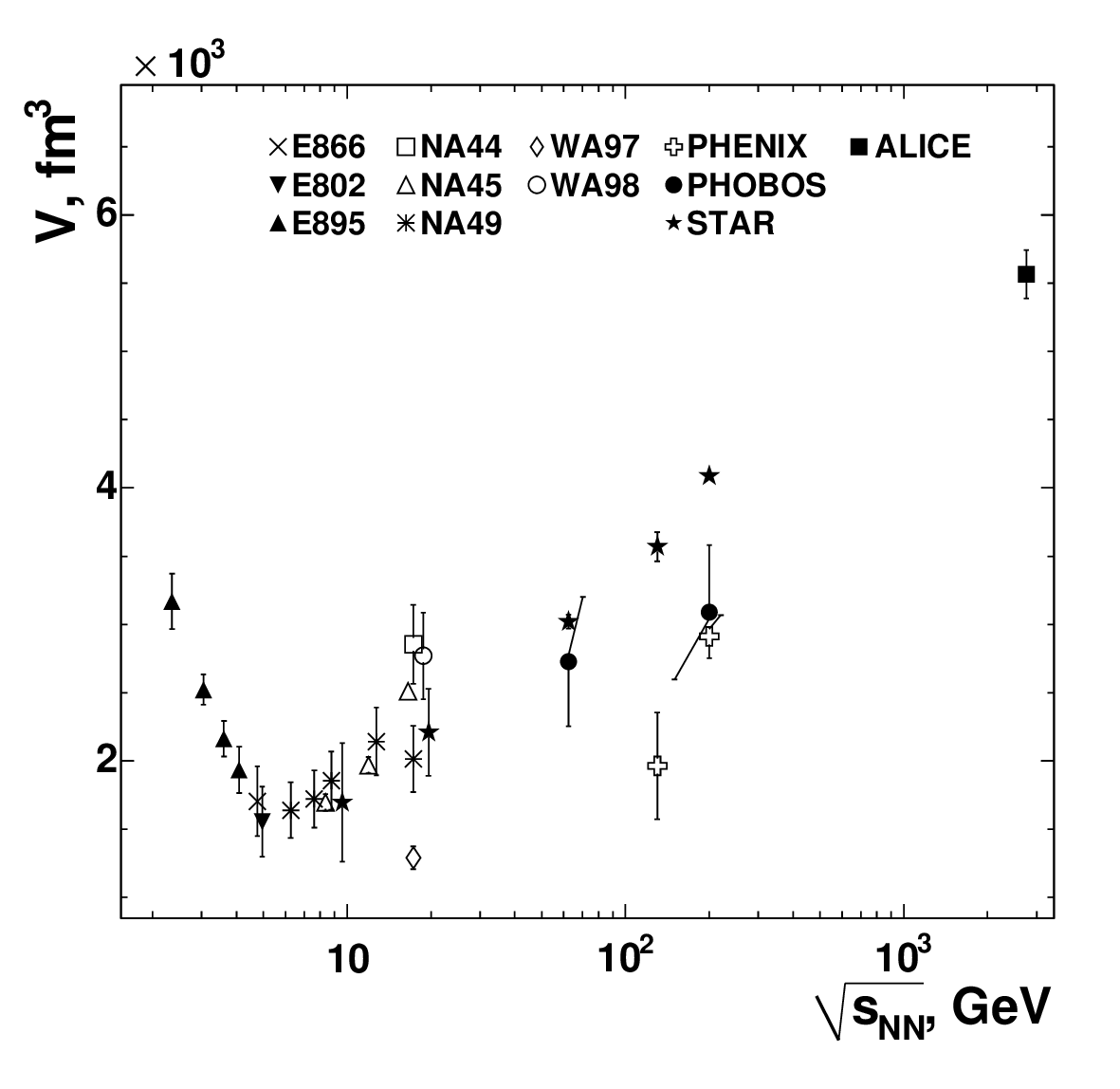}
\vspace*{8pt} \caption{Energy dependence of volume of emission
region at freeze-out for secondary charged pions in central heavy
ion collisions $\mbox{Au+Au, Au+Pb, Pb+Pb}$ in midrapidity region
and at $\langle k_{\perp}\rangle \simeq 0.2$ GeV/$c$. Experimental
results are shown for the same particle types and Coulomb
corrections as well as in Fig.\,\ref{fig:1}. Error bars are only
statistical (for NA44 -- total uncertainties).} \label{fig:2}
\end{figure*}

The volume of homogeneity region in various heavy ion collisions
is calculated based on (\ref{eq:2.4}) and known HBT-radii which
are shown in Figs. \ref{fig:1}b -- d. The pion pairs with low
$k_{\perp}$ are used in these calculations. Thus the homogeneity
region volume obtained for such pairs can be considered as the
estimation of volume of all emissions region. The energy
dependence of estimations of emission region volume is shown in
Fig. \ref{fig:2}. As seen from the figure, the increasing of $V$
with growth of collision energy starts with
$\sqrt{\smash[b]{s_{\footnotesize{NN}}}} \simeq 5$ and it is close
to the (quasi)linear behavior with $\ln(s/s_{0})$, $s_{0}=1$
GeV$^{2}$. It should be noted that the similar functional
dependence was observed for
$v_{2}(\sqrt{\smash[b]{s_{\footnotesize{NN}}}})$ for similar
collision energy range, where $v_{2}$ is the elliptic flow
\cite{Okorokov-SPMP-201-2008}. The increasing of both the
HBT-radii and the source volume observed in Figs. \ref{fig:1},
\ref{fig:2} is explained by the growth of pion multiplicity for
larger $\sqrt{\smash[b]{s_{\footnotesize{NN}}}}$. The $V$
increases significantly (about 1.5 -- 2 times) at transition from
the largest SPS energy ($\sqrt{\smash[b]{s_{\footnotesize{NN}}}} =
17.3$ GeV) to the highest RHIC energy for heavy-ion mode
($\sqrt{\smash[b]{s_{\footnotesize{NN}}}} = 200$ GeV). On the
other hand linear sizes of pion source change weaker in this range
of $\sqrt{\smash[b]{s_{\footnotesize{NN}}}} = 200$ (Figs.
\ref{fig:1}b -- d). The HBT-radii, especially,
$R_{\mbox{\scriptsize{l}}}$ increase substantially with growth of
collision energy from RHIC to LHC. Perhaps, this behavior of
$\mathcal{G}_{i}$, $i=2-4$ parameters can be explained as follows.
The absolute increasing of
$\sqrt{\smash[b]{s_{\footnotesize{NN}}}}$ for transition from SPS
to RHIC is significantly smaller than that for further change from
RHIC to LHC. Thus in the first case the energy range under
consideration in not enough for substantial increasing of radii of
pion source. On the other hand the increasing of
$\sqrt{\smash[b]{s_{\footnotesize{NN}}}}$ on $\approx 2.5$ TeV in
the last case leads to clear growth all geometric parameters of
emission region.

\begin{figure*}
\includegraphics[width=15.5cm,height=17.0cm]{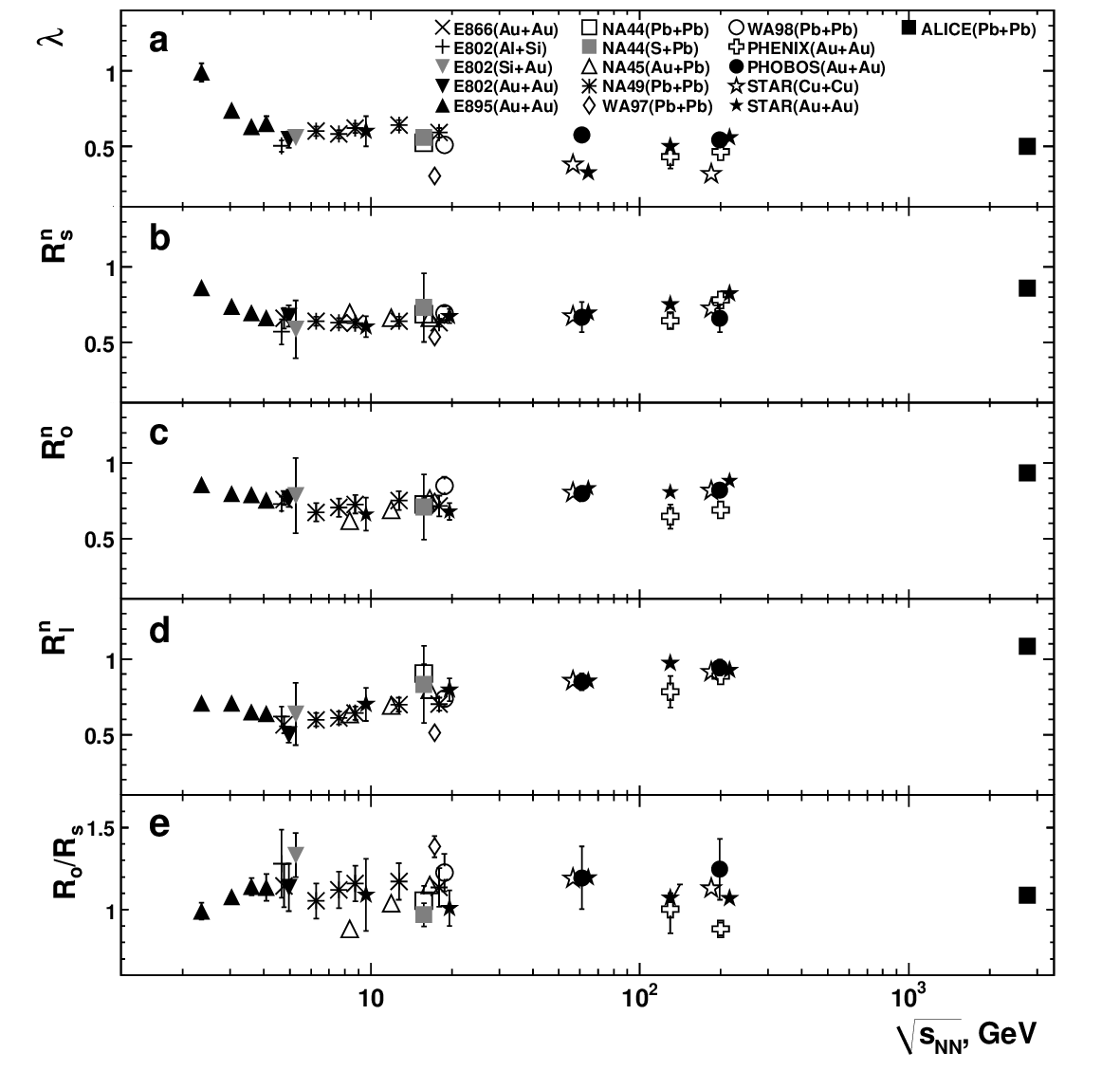}
\vspace*{8pt} \caption{Energy dependence of $\lambda$ parameter
(a), normalized HBT-radii (b -- d) and ratio
$R_{\mbox{\scriptsize{o}}}/R_{\mbox{\scriptsize{s}}}$ (e) in
various nucleus-nucleus collisions at $\langle k_{\perp}\rangle
\simeq 0.2$ GeV/$c$
\cite{Adler-PRL-87-082301-2001,Okorokov-PRC-71-044906-2005,PHENIX-PRL-93-152302-2004,
Ahle-PRC-66-054906-2002,Soltz-NPA-661-439c-1999,Bearden-PRC-58-1656-1998}.
Experimental results are shown for central collisions (for minimum
bias event in the case of E802 for $\mbox{Al+Si}$), for pairs of
$\pi^{-}$ mesons (in cases ALICE and STAR for $\mbox{Cu+Cu}$ --
for $\pi^{\pm}\pi^{\pm}$ pairs, E802 for $\mbox{Al+Si}$, NA44 for
$\mbox{S+Pb}$ -- for pairs of $\pi^{+}$ mesons) and for standard
Coulomb correction $P_{\mbox{\scriptsize{C}}}^{(1)}(q)$ (in cases
ALICE, NA44, NA45, PHOBOS, STAR both for $\mbox{Cu+Cu}$ and for
$\mbox{Au+Au}$ at $\sqrt{\smash[b]{s_{\footnotesize{NN}}}}=62.4$
GeV -- for correction $P_{\mbox{\scriptsize{C}}}^{(3)}$).
Statistical errors are shown (for NA44 -- total uncertainties).}
\label{fig:3}
\end{figure*}

\begin{figure*}
\includegraphics[width=15.5cm,height=17.0cm]{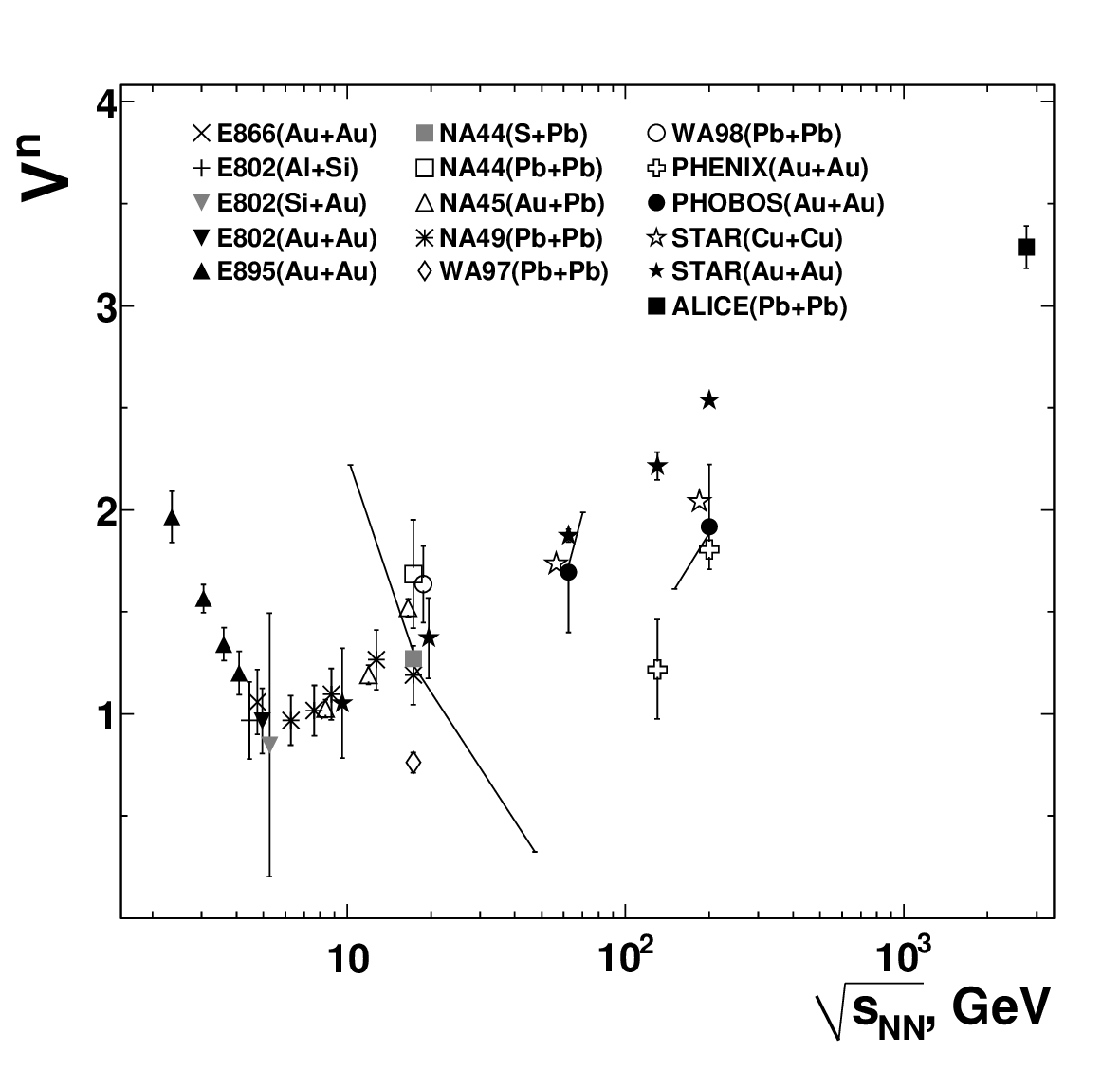}
\vspace*{8pt} \caption{Energy dependence of normalized volume of
emission region at freeze-out for secondary charged pions in
various nucleus-nucleus collisions at $\langle k_{\perp}\rangle
\simeq 0.2$ GeV/$c$. Experimental results are shown for the same
collision, particle and Coulomb correction types as well as in
Fig.\,\ref{fig:3}. Error bars are only statistical (for NA44 --
total uncertainties).} \label{fig:4}
\end{figure*}

Fig. \ref{fig:3} shows the energy dependence of $\lambda$ (a),
normalized HBT-radii (b -- d) and
$R_{\mbox{\scriptsize{o}}}/R_{\mbox{\scriptsize{s}}}$ ratio (e)
for both the symmetrical and the non-symmetrical collisions of
various nuclei. The corresponding dependence for $V^{n}$ is
demonstrated in Fig. \ref{fig:4}. As usual the femtoscopic
parameters from the set $\mathcal{G}$ depend on sign of electrical
charge of secondary pions weakly. Thus the results for
$\pi^{+}\pi^{+}$ pairs obtained in the experiments E802 for
$\mbox{Al+Si}$ collisions \cite{Ahle-PRC-66-054906-2002} and NA44
for $\mbox{S+Pb}$ \cite{Bearden-PRC-58-1656-1998} collisions are
shown in Figs. \ref{fig:3} and \ref{fig:4} also. As seen these
results are in a good agreement with common trends. The large
errors in Fig. \ref{fig:4} for strongly non-symmetrical nuclear
collisions is dominated by large difference of radii of colliding
moderate and heavy nuclei and corresponding large uncertainty for
$\langle R_{\mbox{\scriptsize{A}}}\rangle$. The energy
dependencies for set $\mathcal{G}$ of femtoscopic parameters shown
in Figs. \ref{fig:3} and \ref{fig:4} demonstrate the reasonable
agreement between the values of normalized parameters
(\ref{eq:2.8}) obtained for (quasi)symmetrical collisions of
moderate nuclei and for strongly asymmetrical nuclear-nuclear
interactions $\mbox{Si+Au}$, $\mbox{S+Pb}$ with results for
(quasi)symmetrical heavy ion collisions. Therefore the method
suggested in the paper for normalized femtoscopic parameters
allows us to unite the study both symmetrical and non-symmetrical
nuclear collisions in the framework of general approach. It seems
this approach allows us to obtain the general energy dependencies
of femtoscopic parameters for both the nucleus-nucleus and the
proton-(anti)proton collisions. This investigation is in the
progress at the present time.

\section{Generalized parameterization for the correlation function}\label{sec:4}
The shape of peak of the correlation function contains the unique
experimental information about space-time structure of secondary
particle source at freeze-out. The some physics investigations
confirm the importance of detail study of shape of two-particle
correlation function (see, for example,
\cite{Okorokov-PRC-71-044906-2005}). The parameterization of
$\mathbf{K}_{2}({\bf A})$ depends on type of distribution which
was chosen for emission region. In general there is rich class of
random processes with additive stochastic variables for which
(i.e. for these processes) there are finite distributions but the
Central Limit Theorem (CLT) in the traditional (Gaussian)
formulation is not valid. The class of random processes under
considered are characterized by large fluctuations, power-law
behavior of distributions in the range of large absolute values of
random variables, non-analytic behavior of characteristic function
of the probability distribution for small values of its arguments
\cite{Csorgo-EPJ-C36-67-2004}. In mathematical statistics and
probability theory the class of such distributions are called as
stable (on L\'{e}vy) distributions\footnote{In literature for
physics and mathematics the multidimensional distributions
included in the class are called as L\'{e}vy\,--\,Feldheim
distributions.} \cite{Feller-book-1967}. The general stable
distribution is described by four parameters: an index of
stability (or L\'{e}vy index) $\alpha \in (0,2]$, a skewness
parameter $\beta$, a scale parameter $\gamma$ and a location
parameter $\delta$. These distributions satisfy with requirements
of generalized Central Limit Theorem (gCLT) and
self-similarity\footnote{The applications of stable distributions
in the physics of fundamental interactions and, in particular, for
correlation femtoscopy are described, for example, in
\cite{Okorokov-UchPosob-2009}.}. Therefore the detail
investigation of the shape of correlation peak have to do with
verification of hypothesis of possible self-affine fractal-like
geometry of emission region. At present the study of
L\'{e}vy\,--\,Feldheim distributions is the advanced region of
mathematics but the specific case of central-symmetrical stable
distributions is known in more detail
\cite{Samorodnitzky-book-1994}. Just this subclass of stable
distributions is most important on the point of view of
correlation femtoscopy. In this case the application of subset of
non-isotropic central-symmetrical L\'{e}vy\,--\,Feldheim
distributions \cite{Uchaikin-ZETF-124-903-2003} seems reasonable
because the projections of $\vec{q}$ are independent random
variables.

In accordance with discussion above the generalization is made in
the paper of the experimental results for $\mbox{Au+Au}$
collisions at $\sqrt{s_{\footnotesize{NN}}}=200$ GeV
\cite{Okorokov-PRC-71-044906-2005} and model-independent approach
for study of shape of correlation peak
\cite{Csorgo-PLB-489-15-2000}. In general case the following
multidimensional phenomenological parameterization of n-th order
for CF (\ref{eq:2.1}) is suggested:
\begin{subequations}
\begin{equation}
\hspace*{-2.3cm}C_{2}^{\,\mbox{\scriptsize{ph}},n}(q,K)=\xi_{1}(q,K)\left[1+
\xi_{2}(q,K)\mathbf{K}_{2}^{\mbox{\scriptsize{ph}},n}({\bf
A})\right], \label{eq:4.1.a}
\end{equation}
\vspace*{-0.5cm}
\begin{equation}
\mathbf{K}_{2}^{\mbox{\scriptsize{ph}},n}({\bf A})=\mathbf{K}_{2}
^{\mbox{\scriptsize{ph}},0}({\bf A})
\prod\limits_{i=1}^{3}\prod\limits_{j=1}^{3}
\biggl[1+\sum\limits_{m=1}^{n}g_{m}h_{m}
(A_{ij})\biggr],~~\mbox{at}~n \geq 1. \label{eq:4.1.b}
\end{equation}
\label{eq:4.1}
\end{subequations}
\hspace*{-0.2cm}where $\mathbf{K}_{2}^{\mbox{\scriptsize{ph}},n}$
-- phenomenological parameterization of n-th order for cumulant
correlation function (\ref{eq:2.2}), functions $\xi_{1,2}(q,K)$
take into account formally all corrections on degree of source
chaoticity, final state interactions, etc. The experimental and
theoretical investigations in the field of correlation femtoscopy
allow us to derive some approach for cumulant two-particle
function (\ref{eq:2.2}) in the lowest order. Within the framework
of the subset of non-isotropic central-symmetrical
L\'{e}vy\,--\,Feldheim distributions the most general
parameterization of $\mathbf{K}_{2}^{\mbox{\scriptsize{ph}},0}$
can be given by
\begin{equation}
\mathbf{K}_{2}^{\mbox{\scriptsize{ph}},0}({\bf
A})=\prod\limits_{i=1}^{3}\prod\limits_{j=1}^{3}
\mathbf{K}_{2}^{\mbox{\scriptsize{ph}},0}(A_{ij})=
\exp\biggl(-\sum\limits_{i,j=1}^{3}|A_{ij}|^{\alpha/2}\biggr),~~
\mathbf{K}_{2}^{\mbox{\scriptsize{ph}},0}(x)=\exp(-|x|^{\alpha/2}),
\label{eq:4.2}
\end{equation}
Here were take into account that $x \equiv (q_{i}R_{i})^{2}$,
$i=\mbox{l}, \mbox{o}, \mbox{s}$ for correlation femtoscopy, the
products are on the space components of vectors. The
$\left.\{h_{n}(x)\}\right|_{n=0}^{\infty}$ is the closed system of
orthogonal polynomials in the Hilbert space $\mathcal{H}$:
$\displaystyle \int
dx\mathbf{K}_{2}^{\mbox{\scriptsize{Ô}}}(x)h_{n}(x)h_{m}(x)=\delta_{nm}$,
$g_{n}=\displaystyle \int
dx\mathbf{K}_{2}^{\mbox{\scriptsize{Ô}}}(x)h_{n}(x)$. The system
$\left.\{h_{n}(x)\}\right|_{n=0}^{\infty}$ for exponential weight
function can be derived with the help of the following recurrent
relations $a_{1}h_{1}(x)=(x-b_{0})h_{0}(x)$,
$a_{n+1}h_{n+1}(x)=(x-b_{n})h_{n}(x)-a_{n-1}h_{n-1}(x), n=1,2,...$
\cite{Stahl-book-1992} and moments $\displaystyle
\mu_{n}=\int_{-\infty}^{\infty}dx
x^{n}\exp(-|x|^{\gamma})=2\gamma^{-1}\,\Gamma\bigl(\gamma^{-1}[n+1]\bigr),~n
\geq 0,~\gamma > 0$ \cite{Prudnikov-IntegraliBook1-1981}. Here
$\forall~n \geq 0: b_{n}=\tilde{H}_{n+1}H_{n+1}^{-1}-$
$\tilde{H}_{n}H_{n}^{-1}$; $\forall~n
> 0: a_{n}=H_{n}^{-1}\sqrt{\mathstrut H_{n-1}H_{n+1}}$, and
$H_{n}, \tilde{H}_{n}$ are the following determinants:
$$
H_{n}=
\begin{vmatrix}
\mu_{0}   & \dots & \mu_{n-1} \\
\vdots    &       & \vdots \\
\mu_{n-1} & \dots & \mu_{2n-2} \\
\end{vmatrix},~~~
\tilde{H}_{n}=
\begin{vmatrix}
\mu_{0}   & \dots & \mu_{n-2}  & \mu_{n}\\
\vdots    &       & \vdots     & \vdots \\
\mu_{n-1} & \dots & \mu_{2n-3} & \mu_{2n-1}\\
\end{vmatrix},
$$
$H_{0}=1$ and $\tilde{H}_{0}=0$, the $h_{0}(x)=\mbox{const} > 0$
is defined by normalization which is chosen for system
$\left.\{h_{n}(x)\}\right|_{n=0}^{\infty}$ under consideration.
The specific case $\alpha=1$ and $\alpha=2$ correspond to Cauchy
and Gauss distributions respectively which are widely used in the
correlation femtoscopy. For the first case the Laguerre
polynomials, $L_{n}(x)$, are used as
$\left.\{h_{n}(x)\}\right|_{n=0}^{\infty}$; the Hermite
polynomials, $H_{n}(x)$, are chosen as the closed system of
orthogonal polynomials for the second specific case
\cite{Csorgo-PLB-489-15-2000}.

Perhaps, the generalized parameterization (\ref{eq:4.1}) contain
the important physical information concerning the possible high
irregular geometry of emission region and dynamics of its creation
which is additional with respect to information derived for set
$\mathcal{G}$ of femtoscopic parameters based on traditional Gauss
parameterization. It seems the future development of theoretical
formalism is essential to definition of presence of this new
physical information.

\section{Summary}\label{sec:5}
The following conclusions can be obtained by summarizing of the
basic results of the present study.

The dependencies of femtoscopic characteristics of emission region
on $\sqrt{\smash[b]{s_{\footnotesize{NN}}}}$ are studies for
collisions of various ions. These dependencies are obtained for
range of all experimentally available initial energies and for
estimations of femtoscopic parameters from set $\mathcal{G}$
derived in the framework of Gauss approach. For the first time the
normalized values of radii and volume of source are suggested to
use for energy dependence. This suggestion allows us to expand the
set of interaction types, in particular, on non-symmetrical
nucleus-nucleus collisions which can be studied in the framework
of common approach. There are no the sharp changing of femtoscopic
parameter values, in particular,
$R_{\mbox{\scriptsize{o}}}/R_{\mbox{\scriptsize{s}}}$ with
increasing of $\sqrt{\smash[b]{s_{\footnotesize{NN}}}}$ which were
predicted by some phenomenological models as signature of first
order phase transition in strongly interacting matter.

The generalized parameterization for
$C_{2}^{\mbox{\scriptsize{E}}}(q,K)$ is suggested. This
parameterization takes into account the expansion in closed system
of orthogonal polynomials for general case of non-isotropic
central-symmetrical L\'{e}vy\,--\,Feldheim distribution.

\newpage
\newpage
\newpage
\newpage

\end{document}